\documentclass[referee]{cjaa}           
\input{epsf.sty}                        
\input{psfig.sty}                       
 
\begin{document} 
 
   \title{ Variability of Soft X-ray Spectral Shape in Blazars Observed by ROSAT} 
 
\author{Linpeng Cheng \mailto{}, Yongheng Zhao and Jianyan Wei
     }

   \offprints{L. P. Cheng}

   \institute{National Astronomical Observatories, Chinese Academy of Sciences, Beijing 
100012, P.R. China\\
 \email{clp@lamost.bao.ac.cn}
 }
   
\date{CJAA} 
 
   \abstract{ In paper I (Cheng et al. \cite{Cheng2001}) we have shown that the soft 
X-ray spectra of two types of Seyfert 1 galaxies statistically vary differently 
with increasing intensity. In order to understand how the spectrum of blazars changes, 
the spectral shape variability of 18 blazars observed by ROSAT/PSPC mode are 
studied by presenting the correlation of Hardness Ratio 1 versus Count Rates (HR1-CTs). 
According to our criteria, 10 blazars show a positive HR1-CTs relation, and only 2 
blazars display an anti-correlation of HR1 versus CTs. The rest 6 blazars do not indicate 
any clear correlation. From these we can see that most blazars of our sample statistically show 
a hardening spectrum during overall flux increase, though some vary randomly. 
By investigating the photon index of these objects and different radiation theories, we argue
 that the dominance of the synchrotron or inverse Compton emission in the soft X-ray band 
may interpret the dichotomy of spectral variability well, and that different spectral variations
might represent a sequence of synchrotron peaked frequency.
   } 
 
   \authorrunning{L. P. Cheng, Y. H. Zhao \& J. Y. Wei}           
   \titlerunning{Spectral Shape Variability in Blazars}   
 
   \maketitle 
%
%
\section{Introduction}           
Blazars, including BL Lac objects, highly polarized and optically violently variable 
quasars, and flat-spectrum radio quasars (FSRQs), are characterized by highly variable non-thermal
 emission which dominates their characteristics from the radio through the $\gamma$-rays. The 
mechanism believed to be responsible for their broadband emission is synchrotron radiation 
followed by inverse Compton (IC) scattering at higher energies (e.g. Blandford \& Konigl 
\cite{Blandford1979}). Relativistic beaming of a jet viewed at very small angles is the most natural 
explanation for the extreme properties of the class, including violent variability (up 
to 1-5 magnitudes in the optical; see Wagner \& Witzel \cite{Wagner1995}), high $\gamma$-ray 
luminosities in some cases (Mukherjee et al. \cite{Mukherjee1997}), superluminal motion 
(Vermeulen \& Cohen \cite{Vermeulen1994}), and high optical and radio polarization, sometimes 
extending up to $\rm 10\%$ (Catanese \& Sambruna \cite{Catanese2000}). In addition, the multiwavelength 
spectra of blazars usually show two peaks. The first one peaks at infrared to X-ray energies and 
is most probably from synchrotron radiation, which originates from electrons in a relativistic 
jet pointing close to 
the line of sight. The second peaks at $\gamma$-ray band from GeV to TeV energies and is 
dominated by inverse Compton emission from low-frequency seed photons (Georganopoulos 
\cite{Georganopoulos2000}), which may be the synchrotron photons themselves (Sychrotron 
self-Compton radiation (SSC)), UV photons coming from a nearby accretion disk or from the 
broad-line region (Sambruna et al. \cite{Sambruna1995}). However, the origin of the high 
energy emission is still a matter of considerable debate (e.g. Buckley \cite{Buckley1998}).

   In X-rays, blazars not only exhibit large amplitude variability, but also show significant 
spectral variations with respect to intensity changes. For example, the spectrum of BL Lac PKS 
2005-489 by ROSAT observation softens with decreasing flux (Sambruna et al. \cite{Sambruna1995}).
 EXOSAT observations also indicate its spectral steepening during flux decrease, a behavior 
often displayed by other X-ray strong BL Lac objects (Sambruna et al. \cite{Sambruna1994a}). What is 
more, a similar X-ray spectral variability trend is that their spectrum becomes harder as 
overall flux increases, especially during a flare state. This trend has been consistently
 found by Chiappeiti et al. (\cite{Chiappeiti1999}), Perlman et al. (\cite{Perlman1999}), 
Brinkmann (\cite{Brinkmann2001}) and Romerto et al. (\cite{Romerto2000}), although in different 
energy bands. A possible explanation based on an inhomogeneous jet model is that spectral 
hardening with rising intensity is caused by either the ejection of particles into the jet 
or particle acceleration, and that the spectral steepening was the result of synchrotron cooling 
(Perlman et al. \cite{Perlman1999}; Sambruna et al. \cite{Sambruna1995}).

   The aim of this paper is to show what spectral variation is typical of blazars and
 to give some discussion or interpretation for their spectral variability. Here we 
present a complete spectral shape variability analysis of blazars observed by ROSAT/PSPC 
through the same analysis method as paper I (Cheng et al. \cite{Cheng2001}, 
hereafter paper I). The observations and data reduction are described briefly in Sec. 2, 
and our results are presented in Sec. 3. In Sec. 4 we discuss possible interpretation for 
different spectral variations.
\label{sect:intro} 

 
\section{The Observations and data reduction} 
\label{sect:Obs} 
All the blazars were observed by ROSAT/PSPC mode during the periods from days to years.
 Besides 9 sources selected by the criteria in paper I, we selected some more blazars, 
including BL Lac objects with optical polarization $\rm < 3\%$ (BLs) and high optical polar
ization ($\rm > 3\%$) blazars (HPs), from cross identification of veron (2000)'s AGN catalogue
 with ROSAT pointed catalog. Applying the ROSAT public archive of PSPC observations, 
the sources with average Count Rates (CTs) more than 0.05 $\rm s^{-1}$ are selected so that 
the error of data points is moderate. This yielded 25 blazars. The datasets 
are then processed for instrumental corrections and background subsection using the 
EXSAS/MIDAS software.

The light curve for each blazar is obtained from original ROSAT datasets with time 
binning of 400 seconds in three energy bands: 0.1-2.4 keV (overall band), 0.1-0.4 keV (
A band), 0.5-2.0 keV (B band). Then we pick up nine of twenty-five objects by the 
following criteria: 1) for each source the ratio of maximal CTs to minimal CTs is greater
 than 2, which assures that the range of CTs variability is large enough; 2) the data 
points are not too scarce ($>5$) and they distribute in one diagram consecutively; 3) HR1 
error is small ($<$ 40$\%$). These sources include five BLs and four HPs. Thus 18 
blazars are included in our sample. 

\section{Results of spectral shape variability Analysis}
All these X-ray count rates in 0.1-2.4 keV band were gained from original ROSAT 
observations with time binning of 400 s. In addition, 4 energy bands are shown: 
A 0.1-0.4 keV, B 0.5-2.0 keV, C 0.5-0.9 keV, D 0.9-2.0 keV. The standard hardness 
ratios, HR1 and HR2, for ROSAT-PSPC data are defined as:
   \begin{equation}
  HR1 = \frac{B-A}{B+A},  HR2 = \frac{C-D}{C+D}
   \end{equation}
In order to describe the spectral shape variability, we present the HR1-CTs correlation for 
18 blazars, as shown in Figure 1, and the results are listed in Table 1. To 
distinguish different variation trend of each object we fit the data through a linear 
formula (HR1=a+b$\times$CTs): when the slope b is a positive or negative value and 
its relative error is less than 50\%, we think it has a positive or negative correlation;
 the other instances are of random or no clear correlation. These correlations are 
summarized as the following:
\begin{enumerate}
      \item For 18 blazars in our sample, ten of them which include 5 BLs and 5 HPs, 
show a positive HR1-CTs correlation in the sense that the spectrum hardens as the overall 
flux increases, in common with most of previous observations in blazars.
      \item There are 6 objects, 3 BLs and 3 HPs, displaying random variation of the 
HR1 versus CTs relation. In other words, their spectra do not exhibit a clear softening
 or hardening trend with increasing flux.
      \item Two exceptional sources, HP S5 1803+78 and possible BL Lac object 1207+39W4, 
indicate an anti-correlation of the HR1 versus CTs, implying that their spectrum 
steepens with rising intensity, a behavior rarely observed in blazars. 
      \item Considering the HPs and BLs separately in Table 1, we can see that the overall 
 photon index decreases from the BLs with a positive HR1-CTs correlation to those BLs showing
 random relation of HR1-CTs. On the other hand, the HPs show an opposite trend of the photon
 index change to the BLs: the HPs with a positive correlation to those displaying random
 HR1-CTS correlation exhibit a sequence of increasing soft X-ray slope. The average photon
 index of different subgroups is 2.70$\pm$0.21, 2.38$\pm$0.40, 2.00$\pm$0.23, 2.58$\pm$0.37 
 for the BLs with a positive HR1-CTs and random correlation, and the HPs showing a positive and random 
 relation of HR1 versus CTs, respectively. It appears that the two groups, HPs and BLs, 
 though attributed to the same class blazars, behave differently. 
\end{enumerate}
 
 \begin{table*}
  \caption{Spectral shape variation of our selected blazars. (Positive: the HR1-CTS 
 relation is positive, Negative: the HR1-CTS relation is 
 negative, None: the HR1-CTS relation is random; HP: High Optical Polarization blazars
 ($> \rm 3\%$), BL: BL Lac objects with optical polarization $< \rm 3\%$, BL?: a possible 
  BL Lac object; $\rm \Gamma_{rosat}$: the fitted photon index by a power law with a free 
  neutral absorption)  }
  \label{Tab:publ-works} 
  \scriptsize
  \begin{center}
  \begin{tabular}{rllccccccccc} 
  \hline
Name & ROSAT name & RA & DEC & z & Type & $\rm \Gamma_{rosat}$ &  HR1-CTs \\
  & ( 1RXPJ)  &(2000)&(2000)& & & & correlation\\ \hline
 RX J0916$+$52 & 091648$+$5239.3 & 09 16 53.5 & 52 38 28 & 0.190 & BL & 2.82 & Positive \\
 1E S1212$+$078 & 121510$+$0732.0 & 12 15 10.9 & 07 32 02 & 0.136 & BL & 2.61 & Positive \\
 PKS 2005$-$489 & 200924$-$4849.7 & 20 09 24.8 & $-$48 49 45 & 0.071 & BL & 2.92 & Positive \\
 MS 03313$-$3629 & 033312$-$3619.8 & 03 33 12.3 & $-$36 19 50 & 0.308 & BL & 2.38 & Positive \\
 S5 0716$+$71 & 072152$+$7120.4 & 08 41 24.4 & 70 53 41 & 0.000 & BL &  2.77 & Positive \\
 MS 1332$-$2935 &133531$-$2950.5 & 13 35 30.3 & $-$29 50 42 & 0.250 & BL & 2.10 & None  \\
 2E 0336$-$2453 & 033813$-$2443.6 & 03 38 13.2 & $-$24 43 42 & 0.251 & BL & 2.21 & None\\
 1631.9$+$3719 & 163338$+$3713.3 & 16 33 38.2 & 37 13 13 & 0.115 & BL & 2.84 & None  \\
 1207$+$39W4 & 121026$+$3929.0  & 12 10 26.7 & 39 29 10 & 0.610 & BL? & 2.11 & Negative \\
 PG 1218$+$304 & 122120$+$3010.1 &  12 21 20.7 & 30 10 10 & 0.182 & HP  & 2.21& Positive  \\
 3A 1218$+$303& 122122$+$3010.5 & 12 21 21.9 & 30 10 36 & 0.000 & HP  & 2.28& Positive  \\
 1E1552$+$2020 & 155424$+$2011.2  & 15 54 24.6 & 20 11 47 & 0.222 & HP & 1.89 & Positive \\
 3C 454.3 & 225357$+$1608.7 &  22 53 57.6 & 16 08 53 & 0.859 & HP &  1.73& Positive \\
 3C 345.0 & 164258$+$3948.5 &  16 42 58.7 & 39 48 37 & 0.594 & HP & 1.89 & Positive \\
 B2 1215$+$30 & 121752$+$3006.7 & 12 17 52.1 & 30 07 00 & 0.000 & HP & 3.00 &  None \\
 MS 12218$+$2452 &122422$+$2436.1 & 12 24 22.9 & 24 36 11 & 0.218 & HP & 2.46 & None \\
 2E 1415$+$2557 & 141757$+$2543.5 & 14 17 57.5 & 25 43 35 & 0.237 & HP & 2.2 & None \\
 S5 1803$+$78 & 180042$+$7827.9 & 18 00 42.4 & 78 27 57 & 0.680 & HP & 2.26& Negative \\  

 \hline 
 \end{tabular}
 \end{center} 
 \normalsize 
 \end{table*}

\section{DISCUSSION AND CONCLUSIONS}
The variation in the spectral index during the overall flux change can provide 
insights into the emission mechanism and physical conditions of the group blazars.
 As found previously, BL Lacs show a general hardening of the spectrum during their
 flares and a spectral steepening with fading intensity (Perlman et al. \cite{Perlman1999}; 
Sambruna et al \cite{Sambruna1995}). Instead of soft X-ray photon index, here we have 
presented the correlation of hardness ratio versus count rates to describe the spectral 
variability. Among our sample, ten of eighteen blazars show a hardening spectrum during 
total flux enhance and 6 objects do not exhibit evident spectral variance trend or have 
random variations. The only two exceptions, 1207+39W4 and S5 1803+78, soften with 
increasing flux. These results are consistent with what have been described above. 
The fact that two particular objects indicate softening spectrum during the intensity increase 
was also observed in PKS 2155-304 by Sembay et al. (\cite{Sembay1993}). Next we will discuss 
the implications and possible interpretations with respect to different spectral variations.

There is a general consensus that the multifrequency continuum from blazars at least 
up to the UV band is due to synchrotron radiation from high-energy electrons within a
 relativistic jet (e.g. Konigl 1989). The "curved" shape of the continuum may be due 
to the superposition of different emission regions with different particle spectra 
(inhomogeneous models), or to curvature of the particle spectrum within a single 
emission location or both (Ghisellini, Maraschi \& Treves \cite{Ghisellini1985}). On 
the other hand, the $\gamma$-ray band is widely accepted to come from inverse Compton scattering 
emission based on inhomogeneous or homogenous models (Georganopoulos \cite{Georganopoulos2000}).
 In terms of the X-ray band, two different radiation components share and the relative contribution 
varies with various blazars and different energy states (Cappi et al. \cite{Cappi1994}). 

As stated above, most objects in our sample exhibit a hardening soft X-ray spectrum 
with increasing intensity, a behavior often displayed in other X-ray band. That 
might be a typical feature of the class blazars. In the framework of the inhomogeneous 
SSC model, Sambruna et al. (\cite{Sambruna1995}) have given a good fit to the broadband
 energy distribution of the normal BL Lac object PKS 2005-489 in its high and low state,
 respectively. In Fig. 5 of Sambruna et al. (\cite{Sambruna1995}) it is evident that the
 soft X-ray spectrum could be fitted well by a single synchrotron emission and becomes
 steeper in the low state than in the high state. In addition, the 
similar spectral flattening with increasing intensity can be well fitted and explained by a single IC 
radiation (Madejeski et al. \cite{Madejeski1999}). From these, we can see that the X-ray energy 
spectrum of BL Lac objects consistently becomes harder during intensity rise when 
the energy band is dominated either by a single synchrotron emission or by IC radiation,
 which exactly explained the spectral hardening during the overall flux increase. 
Assuming the model applicable to other blazars, the main spectral variation in this 
paper could be well interpreted. At the same time, it suggest that the variable slope 
and flux of the X-rays may be due to a change in the electron distribution function in 
the inner part of the jet. Possible mechanism to change electron distribution is the 
injection of particle into the jet or in situ particle acceleration.

Besides, there are 6 objects, which show random spectral variability in the sense 
that the spectrum does not indicate a clear change trend with varying intensity. Two 
possible explanations have been proposed. The first one is that the observational span 
time is not suitable, which would constrain the flux and spectral variation analysis. 
The more probable interpretation is that the soft X-ray energy distribution of these 
blazars is shared by two radiative components, the synchrotron and inverse Compton 
emissions. According to radiation theories of blazars, two components can present 
different spectral steepness and variation with changing flux, and thus the blended spectrum might 
display a complex spectral variability though the overall intensity increases 
significantly. To our surprise, two particular objects in our sample showed a spectral 
steepening with rising intensity, a phenomenon rarely observed in blazars. Up to now 
there are only a few cases: spectral softening during flux increase has been seen 
twice in PKS 2155-304 (Sembay et al. \cite{Sembay1993}); S5 0716+714 (Giommi et al. 
\cite{Giommi1999b}) and AO 
0235+164 (Madejeski et al. \cite{Madejeski1996}) exhibited X-ray spectral steepening in their flare 
states. As indicated in Table 1, one of the two blazars, 1207+39W4, is still a possible
 BL Lac object and further identification should determine if it is a peculiar blazar. 
The rest one S5 1803+78, a high optical polarization source, displays the spectral 
variation similar to the intermediate-energy peaked BL Lacs (IBLs) S5 0716+714 and AO
 0235+164. Correspondingly, as described by Perlman et al. (\cite{Perlman1999}) the 
spectral steepening is probably because the X-ray spectrum is dominated by the 
very flat inverse Compton scattering radiation in the low state, but by the soft "tail" 
of the steep synchrotron emission in the high state.

 It is interesting to note that the BLs displaying a steepening spectrum with increasing 
flux statistically have a steeper soft X-ray spectrum than those showing no clear 
spectral variation trend while the HPs indicate a trend opposite to the BLs. For 
blazars it is well accepted that the slope from the synchrotron radiation is much 
steeper than that of the IC emission, and that the X-ray energy distribution of high 
energy-peaked blazars is dominated by the synchrotron emission while the IC radiation 
prepondered the energy band for low frequency-peaked blazars (Perlman et al. \cite
{Perlman1999}). Moreover, it is revealed that the blazars with low optical polarization 
generally show relatively high peaked frequency (Scarpa \& Falomo \cite{Scarpa1997}; 
Padovani \& Gliommi \cite{Padovani1996}). As described above, the slope difference 
between the BLs and HPs could be interpreted below: the soft X-ray spectra of the BLs 
with a hardening spectrum during the overall intensity enhances are dominated by steep 
synchrotron radiation, in contrast, those of the HPs indicating the same spectral 
flattening are mainly attributed to relatively flat IC emission; for the BLs and HPs 
showing random spectral variations and a softening spectrum with rising intensity, the 
energy band at 0.1 $<$ E $<$ 2.4 keV should be dominated by the combination of the 
synchrotron and IC radiation. Thus, the photon index of the BLs varies differently 
from that of the HPs from the objects exhibiting a hardening spectrum to random 
spectral variation during overall intensity increase. Further broad-band energy 
distribution analysis would give a detailed description to the two dichotomous properties.

From the discussions above, it appears that the three groups of blazars represent 
relative dominance of the synchrotron and IC radiation in the soft X-ray band. The 
BLs exhibiting a positive HR1-CTs correlation in our sample may be of the synchrotron
 emission preponderance and usually high-energy peaked blazars. On the contrary, the 
spectrum of those HPs showing an positive correlation of the HR1 versus CTs could be 
dominated by the flat IC radiation and they might be of low-energy peaked blazars. 
What is more, the soft X-ray energy distribution of the blazars whose spectrum varies 
randomly with rising flux is probably dominated by the mergence of the synchrotron and
 IC radiation, while for softening spectra that is induced by an alternation between the 
synchrotron and the IC radiation and their synchrotron peaked frequency would be 
intermediate. Consequently, it seems that from the BLs indicating a hardening spectrum 
through the blazars exhibiting random spectral variation or a spectral softening to 
the HPs showing a hardening spectrum, their synchrotron peaked frequency shifts from 
high to low.

 In conclusion, in this paper we analyzed a complete spectral shape variability of blazars 
by ROSAT/PSPC observations. Most of the blazars in our sample exhibit a hardening 
soft X-ray spectrum with increasing flux, a typical behavior of the subclass; there 
are also 6 blazars which do not show any clear spectral variation trend and 2 
objects display a steepening spectrum with rising intensity, a behavior rarely 
revealed in blazars. Based on the properties of the synchrotron and IC emissions we
 argue that different spectral variations might represent the relative contribution 
of the two components: the soft X-ray spectrum of the BLs with a hardening spectrum 
are dominated by the synchrotron emission while for the HPs it is dominated by the 
IC radiation instead; those showing random spectral variation are prepondered by 
the combination of the two radiations, and the steepening are prevailed by the alternation 
between the IC and synchrotron emissions. Thus, different soft X-ray spectral variations might
 correspond to a sequence of shifting synchrotron peaked frequency. 

\begin{acknowledgements} 
 We thank Dr. Luo Ali and other members of LAMOST Group for helping us with data 
reduction and some software applications. Supports under Chinese NSF (19973014), 
the Pandeng Project, and the 973 Project (NKBRSF G19990754) are also gratefully 
acknowledged.
\end{acknowledgements}

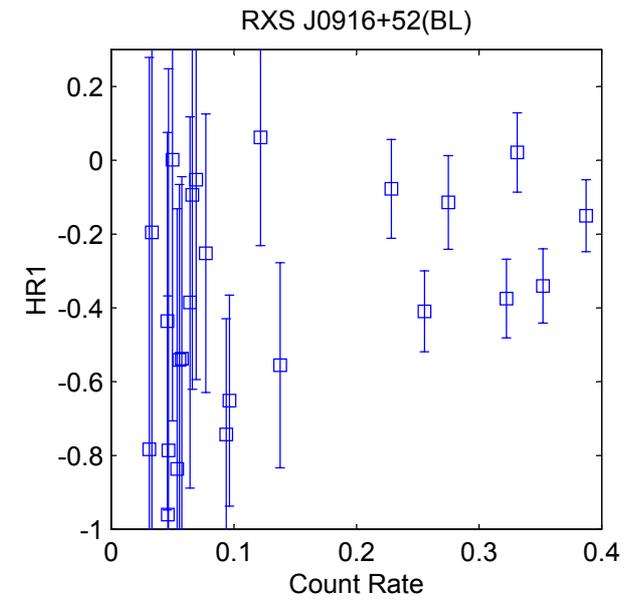
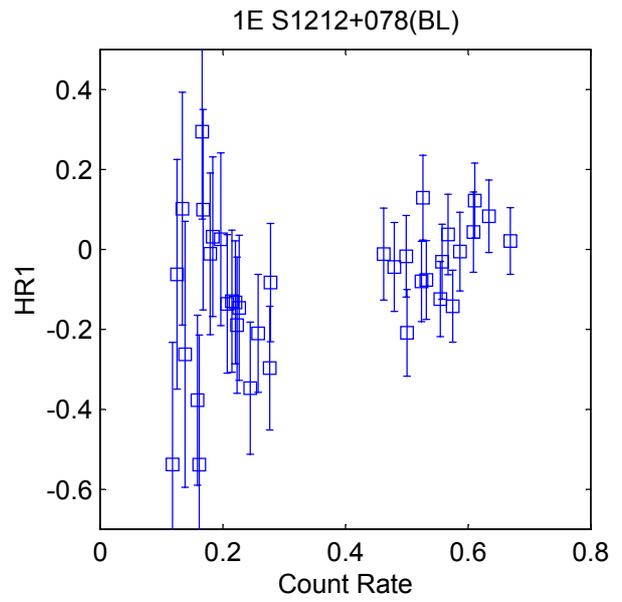
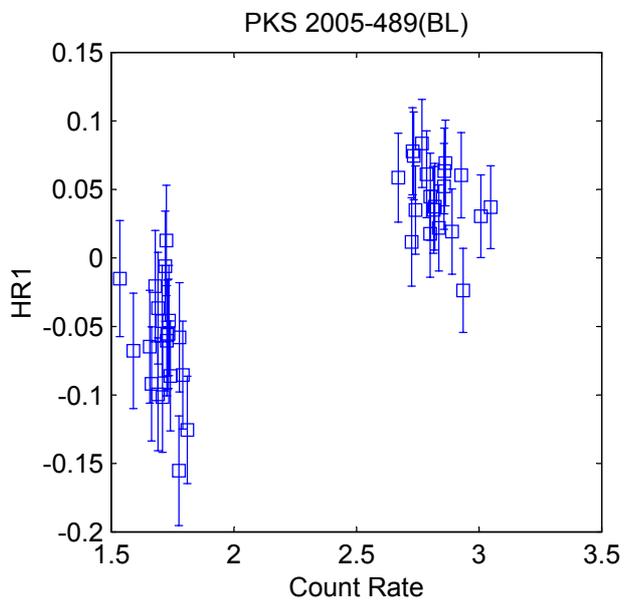
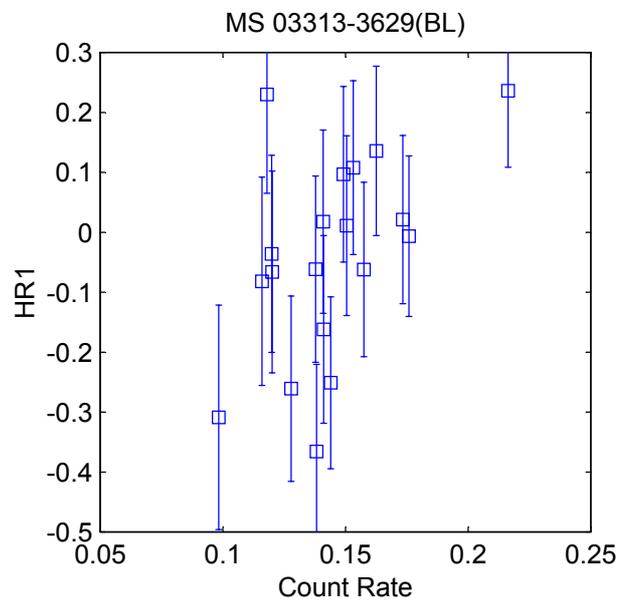
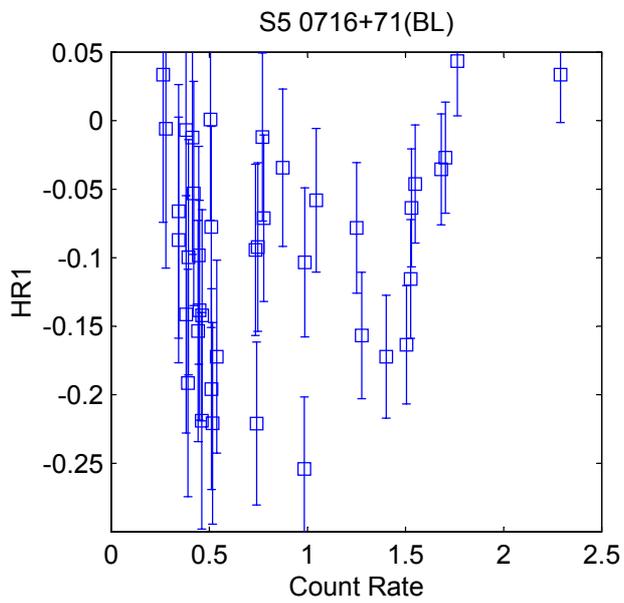
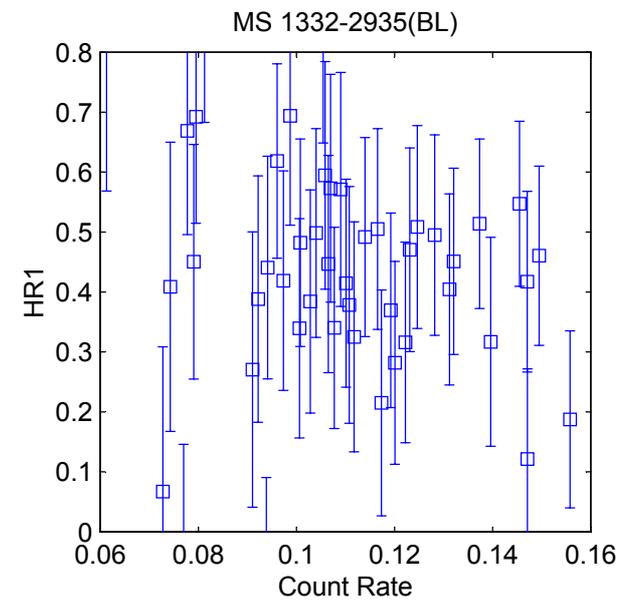
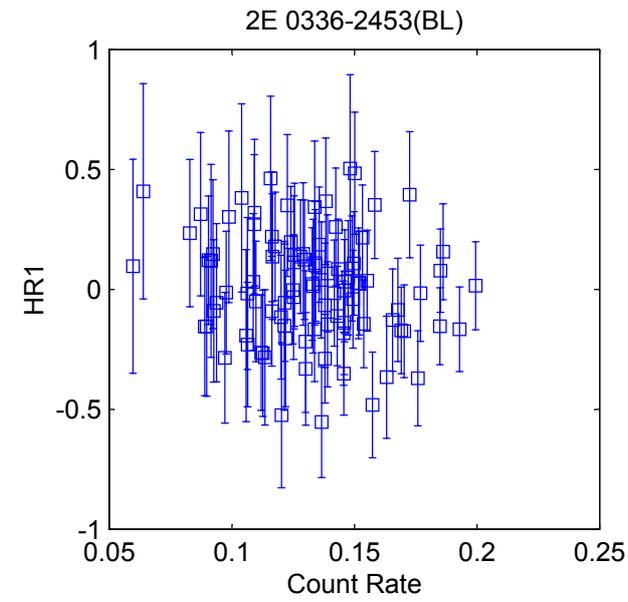
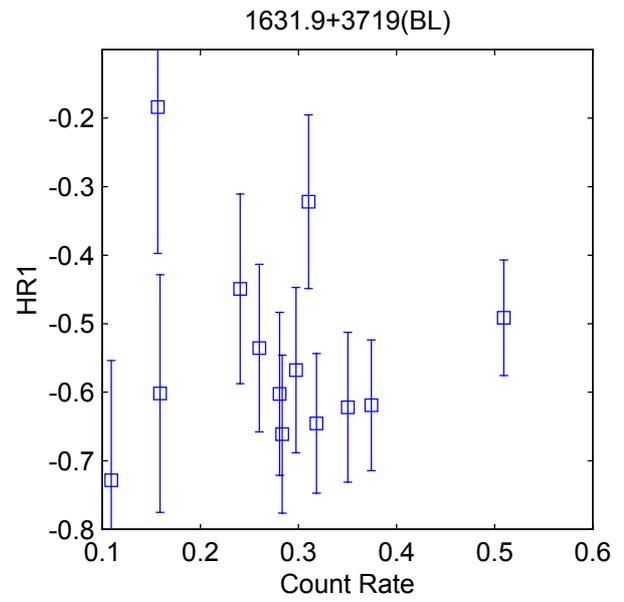
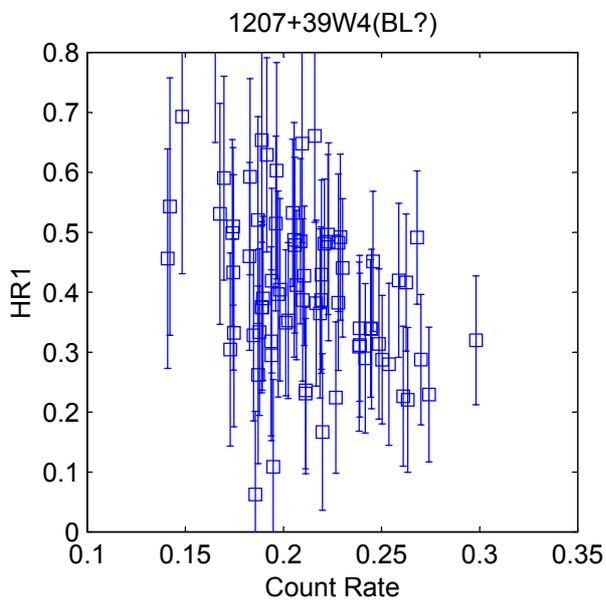
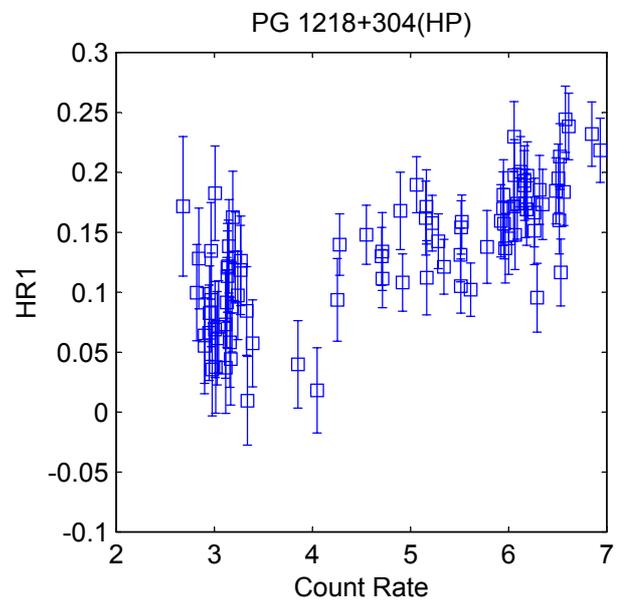
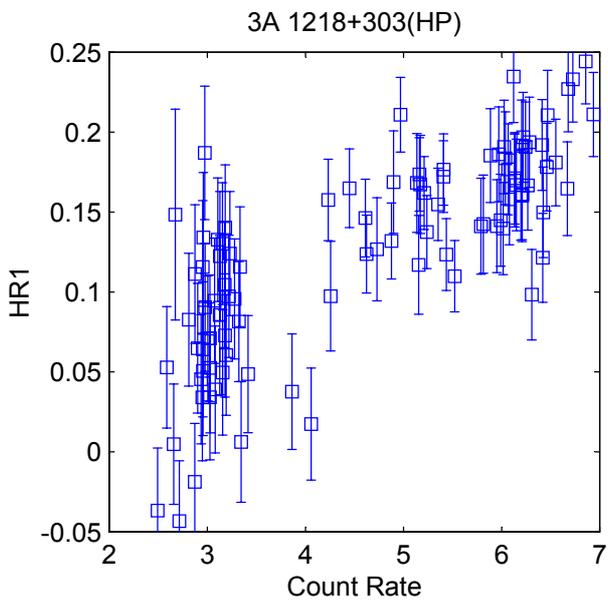
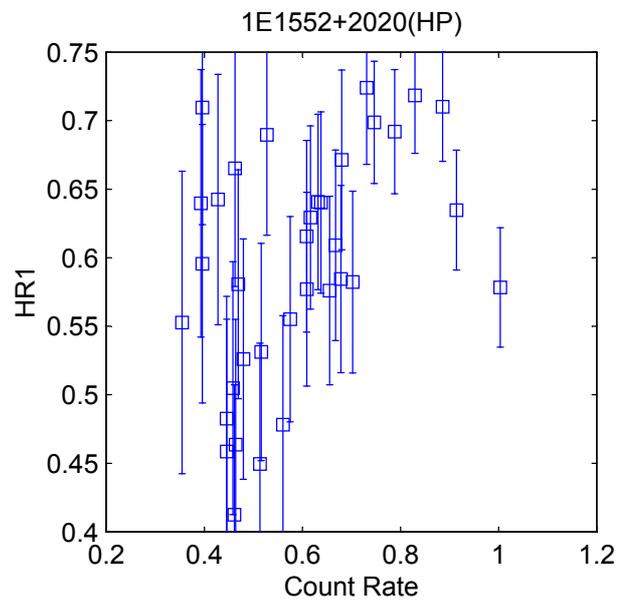
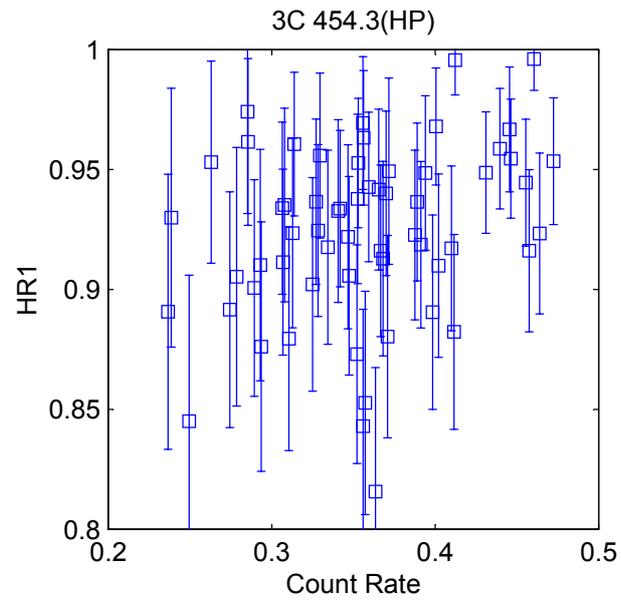
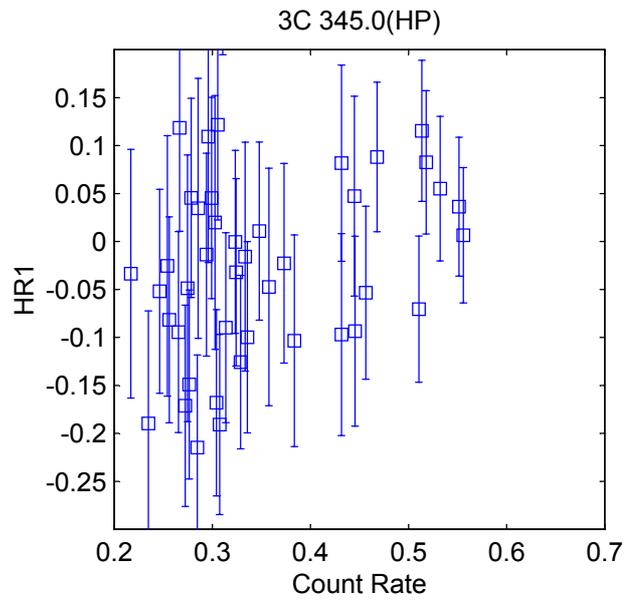
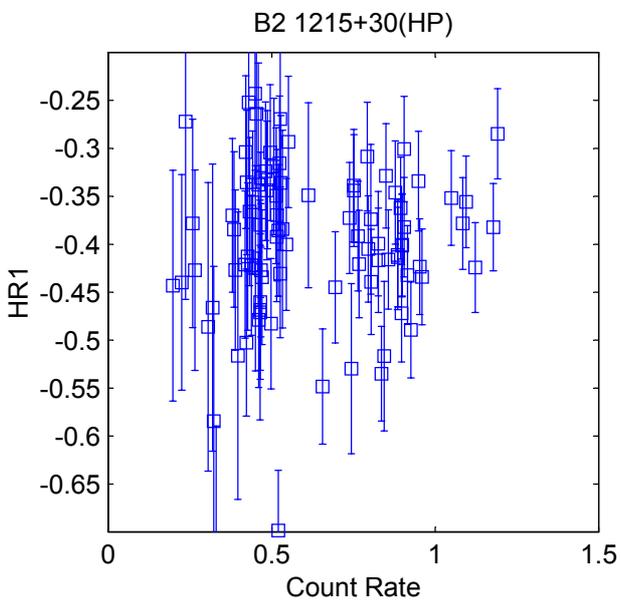
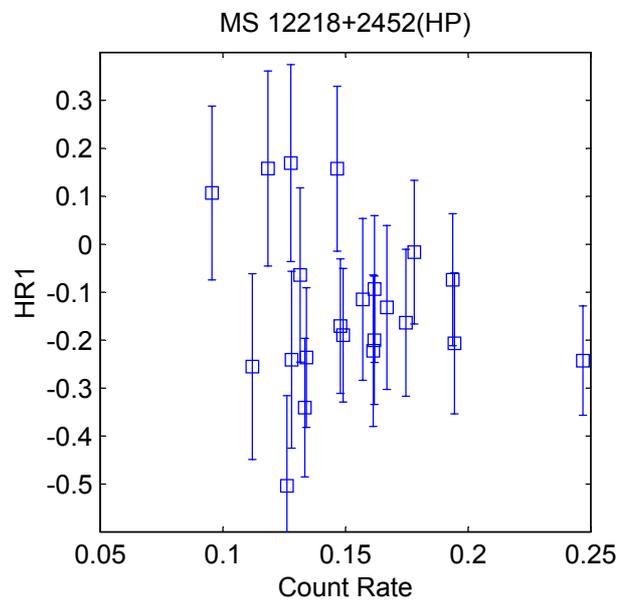
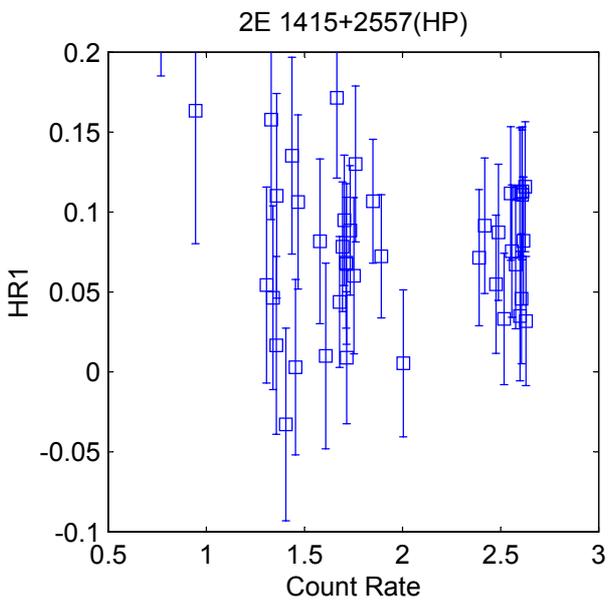
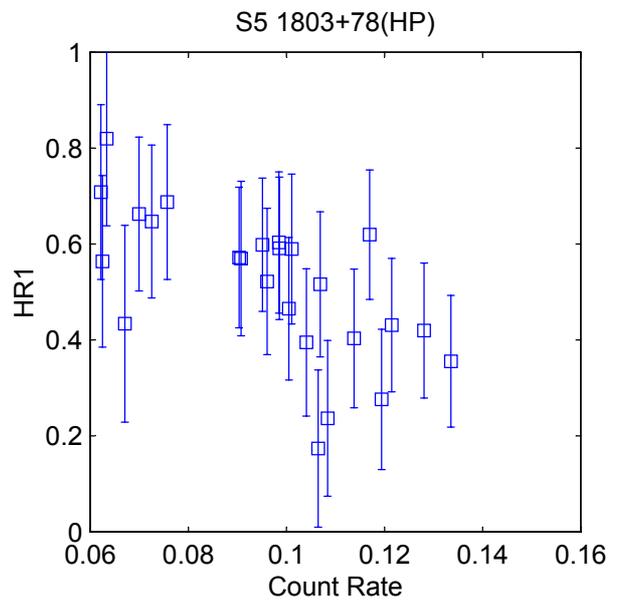
\end{document}